\documentclass[twocolumn,aps,prb,tightenlines,amsmath,amssymb,superscriptaddress]{revtex4}
\usepackage{graphicx}
\usepackage{dcolumn}
\begin{document}
\hyphenation{spectro-meter}
\hyphenation{micro-scopy}
\hyphenation{image}
\hyphenation{Na-tu-rally}
\title{Raman spectroscopy of the interlayer shear mode in few-layer MoS$_2$ flakes}
\author{G. Plechinger}
\affiliation{Institut f\"ur Experimentelle und Angewandte Physik,
Universit\"at Regensburg, D-93040 Regensburg, Germany}
\author{S.\ Heydrich}
\affiliation{Institut f\"ur Experimentelle und Angewandte Physik,
Universit\"at Regensburg, D-93040 Regensburg, Germany}
\author{J.\ Eroms}
\affiliation{Institut f\"ur Experimentelle und Angewandte Physik,
Universit\"at Regensburg, D-93040 Regensburg, Germany}
\author{D.\ Weiss}
\affiliation{Institut f\"ur Experimentelle und Angewandte Physik,
Universit\"at Regensburg, D-93040 Regensburg, Germany}
\author{C.\ Sch\"uller}
\affiliation{Institut f\"ur Experimentelle und Angewandte Physik,
Universit\"at Regensburg, D-93040 Regensburg, Germany}
\date{\today}
\author{T.\ Korn}
\email{tobias.korn@physik.uni-regensburg.de}
\affiliation{Institut
f\"ur Experimentelle und Angewandte Physik, Universit\"at
Regensburg, D-93040 Regensburg, Germany}
\begin{abstract}
Single- and few-layer MoS$_2$ has recently gained attention as an interesting new material system for opto-electronics.
Here, we report on scanning Raman  measurements on few-layer MoS$_2$ flakes prepared by exfoliation. We observe a Raman mode corresponding to a rigid shearing oscillation of adjacent layers. This mode appears at very low Raman shifts between 20 and 30 cm$^{-1}$. Its position strongly depends on the number of layers, which we independently determine using AFM measurements and investigation of the other characteristic Raman modes. Raman spectroscopy of the shear mode therefore is a useful tool to determine the number of layers for few-layer MoS$_2$ flakes.
\end{abstract}
\maketitle
The tremendous growth of experimental research on graphene in the past few years stems, in part, from the simple exfoliation technique that allows for preparation of single- and few-layer flakes from bulk crystals. This technique is applicable to many layered crystal structures in which the binding energy between adjacent planes is much lower than the binding energy within a plane~\cite{Novoselov26072005}. Among these layered structures, the  dichalcogenide MoS$_2$, which is used commercially, e.g., as a  high-temperature dry lubricant, has attracted a lot of interest. It was recently shown to undergo a transition from indirect to direct-gap semiconductor when its thickness is reduced to a single layer~\cite{Heinz_PRL10,Splen_Nano10}, leading to pronounced photoluminescence. This drastic change of the band structure has been investigated theoretically by a number of groups~\cite{Eriksson09,ellis:261908,Kadantsev2012}, and further calculations suggest the possibility of band structure engineering using strain~\cite{Scalise12}. Low-temperature photoluminescence measurements revealed the presence of impurity-bound excitons in single-layer MoS$_2$ flakes~\cite{Korn11}, which can be suppressed in oxide-covered MoS$_2$~\cite{Plechinger12}. The photocarrier lifetime in single-layer MoS$_2$ is sufficiently short~\cite{Korn11} to make the material interesting for fast photodetectors,  and a MoS$_2$-based phototransistor was reported recently~\cite{Yin11}.   Room-temperature transistor operation with very large on/off ratio has been demonstrated for single MoS$_2$ layers~\cite{Kis_NatNano10}. As in graphene, where research was initially focused on single layers and later expanded to study also bilayers and trilayers due to their different band structure,   few-layer MoS$_2$ flakes may be interesting for their transport properties: very recently,   ambipolar transistor operation was shown in few-layer MoS$_2$~\cite{Zhang_NanoLett12}. Both, chemical exfoliation~\cite{Eda11} and vapor phase growth techniques~\cite{Zhan12}  have been demonstrated for MoS$_2$, indicating the possibility of fabricating large-area thin films neccessary for potential applications.   Similar to graphene, where Raman scattering has been used to determine, e.g., the layer thickness~\cite{Ferrari_PRL06} or the doping type and concentration~\cite{Steffi_APL10}, Raman spectroscopy is a highly useful tool to map MoS$_2$ flakes and to identify single layers~\cite{Heinz_ACSNano10,najmaei:013106}. This is facilitated by the fact that two characteristic Raman modes, A$_{1g}$ (out-of-plane optical vibration of the sulfur atoms) and E$^1_{2g}$ (in-plane optical vibration of Mo and S atoms), show opposite frequency dependence on the number of layers:  the A$_{1g}$ mode increases its frequency with the flake thickness,  while the E$^1_{2g}$ anomalously softens due to increased dielectric screening~\cite{Molina11}, so that the difference of the mode frequencies is characteristic for a certain number of layers.
While many of the  Raman modes of graphene and MoS$_2$, or other dichalcogenides, are very different due to the different crystal structures, two  modes are rather generic for layered crystal structures:

(1) a shear mode, in which adjacent layers rigidly oscillate relative to each other, with the oscillation amplitude lying in the layer plane. This mode was observed for, both, bulk graphite~\cite{Nemanich75}, and bulk MoS$_2$~\cite{Verble1972941}. More recently it was detected in few-layer graphene~\cite{Ferrari_NatMat12}.

(2) a compression mode, in which adjacent layers oscillate rigidly, with the oscillation amplitude perpendicular to the layer plane. This mode was recently observed in few-layer graphene in a double resonant process in combination with an LO phonon~\cite{Maultzsch12}.

Here, we report on scanning Raman scattering  measurements on MoS$_2$ flakes. The MoS$_2$ flakes were prepared using the transparent tape liftoff method well-established for graphene, from natural MoS$_2$. A  p-doped  silicon wafer (specific resistivity $\rho= 0.005$~Ohm~cm, corresponding to degenerate doping) with 300~nm SiO$_2$ layer and lithographically defined metal markers was used as a substrate. After initial characterization with an optical microscope, the samples were analyzed by Raman spectroscopy at room temperature. For this, we utilized a microscope setup, in which a 532~nm cw laser was coupled into a 100x microscope objective, which also collected the scattered light in backscattering geometry. The scattered light was recorded using a triple grating spectrometer equipped with a liquid-nitrogen-cooled charge-coupled device (CCD) sensor. The sample was mounted on a piezo-stepper table and scanned under the microscope. The spatial resolution of this setup is about 500~nm. For some of the Raman measurements, a cross-polarized backscattering geometry was used. Here, a polarizer was placed in front of the spectrometer so that only the scattered light with polarization perpendicular to the laser polarization was coupled into the spectrometer. In this scattering geometry, the spectrally broad background, which stems from inelastic light scattering of free carriers in the heavily p-doped Si substrate~\cite{Cardona80}, is suppressed and allows us to obtain Raman spectra very close to the laser line.

\begin{figure}
\includegraphics*[width=\linewidth]{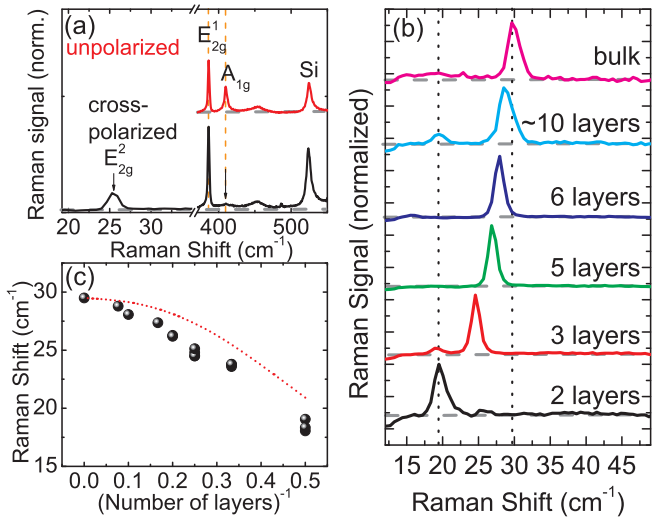}
\caption{(a) Raman spectra of a three-layer MoS$_2$ flake. The top spectrum is recorded without a polarizer in front of the spectrometer, the bottom spectrum in cross-polarized geometry. (b) Normalized Raman spectra of the shear mode measured on MoS$_2$ flakes of different thickness.  (c) Spectral position of the shear mode as a function of (number of layers)$^{-1}$. Multiple datapoints indicate measurements on different flakes. The dotted line indicates  the mode position as predicted by the model developed in ref.~\onlinecite{Ferrari_NatMat12}}
\label{ShearNumber_3panel}
\end{figure}

First, we discuss the typical Raman spectra observed in our samples. Figure~\ref{ShearNumber_3panel} (a) shows two Raman spectra collected on a three-layer area of a flake. The top spectrum is recorded without a polarizer in front of the spectrometer, while the bottom spectrum is recorded in cross-polarized geometry, as described above. Due to the large and spectrally broad background caused by  the silicon free-carrier scattering, the unpolarized spectrum is only recorded in a spectral range starting at 200~cm$^{-1}$ with respect to the laser line, while the cross-polarized spectrum is collected in a range starting at 10~cm$^{-1}$. In the unpolarized spectrum, we note the two characteristic Raman modes E$^1_{2g}$ (386.1~cm$^{-1}$) and A$_{1g}$ (409.7~cm$^{-1}$), with a frequency difference of 23.6~cm$^{-1}$ as expected for three layers~\cite{Heinz_ACSNano10}. By contrast, in the cross-polarized spectrum, the  A$_{1g}$ mode is almost completely suppressed, while the E$^1_{2g}$ mode remains visible. Additionally, in the cross-polarized spectrum, we observe a low-frequency mode at 25.3~cm$^{-1}$, which we associate with the  E$^2_{2g}$ interlayer shear mode. The main focus of this manuscript is the study of its frequency dependence on the number of MoS$_2$ layers.

Our results are summarized in Figure~\ref{ShearNumber_3panel}(b): it shows normalized shear mode spectra obtained from MoS$_2$ flakes of different thickness. The number of layers was determined independently, by, both, atomic force microscopy (AFM) measurements (see below), and investigation of the A$_{1g}$ - E$^1_{2g}$ frequency difference. We clearly observe that the shear mode shifts from below 20~cm$^{-1}$ for a bilayer to about 30~cm$^{-1}$ for bulk-like flakes, a very large relative shift of more than 50~percent. In Figure~\ref{ShearNumber_3panel}(c), we plot the shear mode positions extracted for 16 different flake areas as a function of (number of layers)$^{-1}$. Multiple data points for a given thickness stem from measurements on different flakes. For comparison, we have calculated the expected mode position as a function of (number of layers)$^{-1}$ as predicted in ref.~\onlinecite{Ferrari_NatMat12} using a linear chain model with fixed interlayer coupling strength. Here, we have used our measured value for the shear mode frequency in bulk as the only parameter. We clearly see that our experimental values show a more pronounced shift than expected using the linear chain model. Two factors may contribute to this difference: (a) we have to assume that, since all experiments were carried out under ambient conditions, the topmost layer of the MoS$_2$ flake is partially covered with adsorbates, effectively increasing the mass per unit area. (b) our flakes are not free-standing, but deposited on a SiO$_2$ substrate, therefore, the bottom layer will experience a weak van der Waals interaction with the substrate, which would act as an additional spring constant in the linear chain model.
\begin{figure}
\includegraphics*[width= \linewidth]{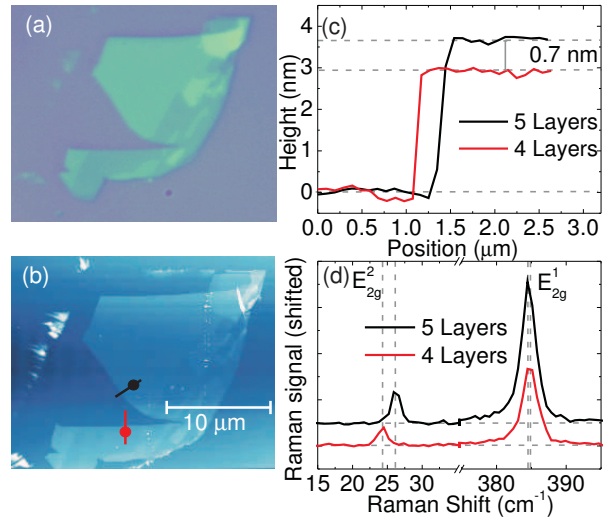}
\caption{(a) Optical micrograph of a few-layer MoS$_2$ flake. (b) AFM image of the same flake. The two lines indicate the areas where the height profiles (shown in (c)) were taken, the dots indicate the points where the Raman spectra (shown in (d)) were recorded.  (c) AFM height profiles for steps from the substrate onto different areas of the flake. (d) Raman spectra measured at two different points on the flake. The dotted lines serve as guide to the eye.}
\label{AFM_Step}
\end{figure}
\begin{figure}
\includegraphics*[width= \linewidth]{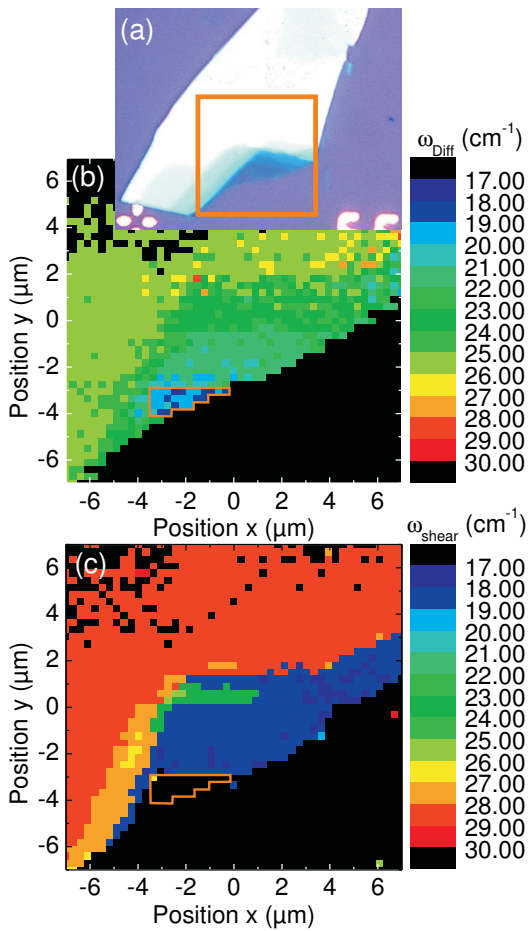}
\caption{(a) Optical micrograph of a few-layer MoS$_2$ flake. The square marks the scan region for the false color plots in (b) and (c). (b) False color plot of the frequency difference between the A$_{1g}$ and E$^1_{2g}$ modes. The outline of the single-layer area of the flake is marked by the solid line. (c) False color plot of the shear mode position. The scan region is identical to (b).}
\label{Shearmapping}
\end{figure}

Figure~\ref{AFM_Step} demonstrates the process of establishing the relation between shear mode position and number of layers in more detail. Utilizing  flakes with large areas of uniform height, determined by optical microscopy, such as the one shown in Fig.~\ref{AFM_Step}(a), we determine the thickness of these areas by extracting height traces  from AFM images. Raman spectra are then recorded at points on the flake for which the height is known. In Fig.~\ref{AFM_Step}(b), the AFM image corresponding to Fig.~\ref{AFM_Step}(a) is shown. The two colored lines indicate the position of the AFM line traces leading from the substrate to the flake shown in Fig.~\ref{AFM_Step}(c). From these traces, using the thickness of a single MoS$_2$ layer of $\backsim$0.7~nm, we find that the black trace corresponds to a step height of 5 layers, while the red trace corresponds to 4 layers. Raman spectra collected on these areas of the flake (Fig.~\ref{AFM_Step}(d)) demonstrate the large blueshift of the shear mode (about 1.8~cm$^{-1}$), which is clearly visible in the data and also easy to determine using an automated fitting routine. By comparison, the E$^1_{2g}$ mode only shows a weak redshift (about 0.5~cm$^{-1}$), so that even though the spectrally integrated intensity of this mode is about 8 times higher than that of the shear mode for a 5-layer region, the shear mode allows for easier discrimination of the number of layers.

The large relative shift of the interlayer shear mode can be used to efficiently map few-layer MoS$_2$ flakes by using the shear mode position. One example is shown in Figure~\ref{Shearmapping}. The few-layer flake shown in Fig.~\ref{Shearmapping}(a) was scanned in the region marked by the square using the Raman microscope with a step size of 300~nm. The collected spectra were analyzed using an automated fit routine, which determines the peak positions of the A$_{1g}$,  E$^1_{2g}$ and E$^2_{2g}$ modes as a function of position. From this data, false color plots were generated. In Fig.~\ref{Shearmapping}(b), we plot the frequency difference A$_{1g}$ - E$^1_{2g}$ as a function of position. From the frequency difference we can clearly identify regions of different thickness, including a small monolayer area, which shows a frequency difference of about 18~cm$^{-1}$ and is marked with a solid outline. Naturally, there is no shear mode signal in this region in  Fig.~\ref{Shearmapping}(c), where we plot the shear mode frequency as a function of position. Comparing the contrast of the two false color images, where we have used the same false color scale, we immediately notice that the shear mode image shows much more well-defined transitions between regions of different thickness and easily allows to identify, e.g., a small trilayer region (green color) within the larger bilayer area (blue color), and a transition from 6 layers (dark yellow color) to about 10 layers (orange color), which are hard to discern from the frequency difference image.

In conclusion, we have used scanning Raman spectroscopy to study the interlayer shear mode in few-layer MoS$_2$ flakes. We observe a large frequency shift of the shear mode to higher energies with increasing number of layers. This large shift allows us to precisely map the layer thickness in few-layer MoS$_2$ flakes.
The authors gratefully acknowledge financial support by the DFG via SFB689, SPP 1285 and GRK 1570 as well as fruitful discussion with L. Wirtz, A. Molina-S\'{a}nchez, J. Maultsch and N. Scheuschner.

\end{document}